\documentclass[12pt]{iopart}
\usepackage{graphicx}

\begin{document}

\title[Contact Resistance in Graphite-Graphene contacts]
      {Electric Contact Resistance in Graphite-Graphene contacts from \textit{ab initio} methods}
\author{Ferran Jovell$^1$, Xavier Cartoix\`a$^1$}
\address{ $^1$ Departament d'Enginyeria Electr\`onica, Universitat Aut\`onoma de Barcelona, 08193 Bellaterra, Spain}
\ead{Ferran.Jovell@uab.cat}

\begin{abstract}
  We study the ballistic transmission and the contact resistance ($R_c$) of a
  graphite-graphene contact in a top contact geometry from first principles.
  We find that the calculated $R_c$'s depend on the amount of
  graphene-graphite overlap, but quickly saturate for transfer lengths of the
  order of 20~\AA. For contacts overlapping more than this transfer length,
  the $R_c$ can be lower than the 100~$\Omega \cdot \mu$m mark. On the other
  hand, edge graphite-graphene contacts are expected to have exceptionally low
  contact resistance.
\end{abstract}

\vspace{2pc}
\\
%

\maketitle

\section{Introduction}
Throughout the last fifteen years, graphene has demonstrated its capabilities as a new
material with its extraordinary properties~\cite{graphenereview}. Although the
lack of bandgap forbids the use of this material for digital applications, its
properties are very well suited for analog radiofrequency devices~\cite{electronic2dreview}.
However, before graphene can be widely adopted, several difficulties must be overcome.
In particular, one of the limitations for the use of graphene in analog electronics is
the high contact resistance when metal-graphene contacts are fabricated, while an
upper bound of $100~\Omega \cdot \mu$m would be desirable~\cite{electronic2dreview,rcbound}.
\\\\
Theoretical work has been carried out for contact resistance between graphene and
other metals. For instance, Chaves \etal~\cite{rcmodel} created a model for
contact resistance between metal and graphene in a top contact-like geometry.
The metal-graphene edge contact geometry, in spite of its vanishing contact
overlap, has also been proven to be at least as good as some of
the top contact geometries~\cite{sidecontact}.

Despite their obvious similar structural properties, the use of graphite as an
electrode for contacting graphene has received much less attention. The Lieber
group has synthesized monolithic graphene-graphite structures, obtaining
specific contact resistivities in the range of 700-900~$\Omega \cdot \mu$m, better than
similarly fabricated Cr/Au junctions~\cite{ggmonolithic}. Also, Chari \etal
measured the resistivity of rotated graphite-graphene contacts, obtaining
specific contact resistivities as low as 133~$\Omega \cdot \mu$m for holes and
200~$\Omega \cdot \mu$m for electrons~\cite{ggcontactexp}.
\\\\
The objective of this paper is to demonstrate the viability of graphite-graphene
contacts and show their fundamental limits. To this purpose, we describe in
\sref{ssec:geometry} the used geometry, followed by the computational
methodology in \sref{ssec:computational}. We show in \sref{sec:results}
that this yields results well below the upper contact resistance limit for certain
values of the overlap area and doping level. Then, an eigenchannel analysis
give us more insight about the scattering processes in the interface between the
graphite substrate and the graphene. This analysis brings us to the conclusions,
in \sref{sec:conclusions}, that the graphite-graphene interface presents more of
an area effect than metal-graphene contacts~\cite{xcs_paper}, but still with very
small transfer lengths of $\sim$20~\AA.

\section{Methodology}
\subsection{Geometry Description \label{ssec:geometry}}

\begin{figure}[t!]
  \centering
  \includegraphics[width=1.0\linewidth]{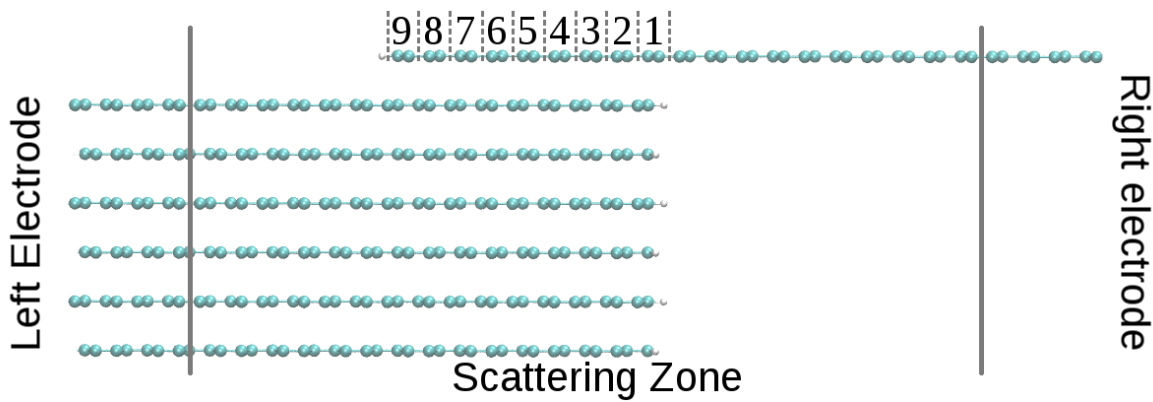}
  \caption{Ball-and-stick structure for the Graphite-Graphene top contact.
  Green/dark gray (white/light gray) balls indicate the carbon (hydrogen) atoms.
  Also indicated are the three different regions: Left Electrode, Scattering Zone
  and Right Electrode used for transport calculations.
  Electrons are injected in the left electrode through the scattering zone into
  the right electrode. }
  \label{fig:geometry}
\end{figure}

We will focus on top contact geometries because they are the most easily fabricated.
In \Fref{fig:geometry}, a ball-and-stick representation of the structure of the
graphite-graphene contact is displayed.
As usual in ballistic transport calculations, there are three differentiated zones:
a left electrode---where the electrons are injected---, a scattering zone through which
electrons will pass or reflect, and a right electrode into which electrons
that were not backscattered will arrive. The electrodes are semi-infinite and we will
be studying a single graphite-graphene contact.

In this work we studied the effect of overlap length of graphene over the graphite
bulk.  The numbers in the scattering zone indicate
the different number of overlapping C-pairs providing the contact between graphene
and the graphite substrate. We also studied the case where the last graphite
layer turns into the graphene sheet, which we interpret as an edge
graphite-graphene contact~\cite{sidecontact}.

Structures were relaxed from first-principles using the {\sc Siesta} code~\cite{siesta},
an efficient implementation of the Density Functional Theory (DFT) using localized pseudo-atomic
orbitals. Transport calculations were carried out using {\sc TranSiesta}~\cite{transiesta,transiesta2},
which implements the Non-Equilibrium Green's Function formalism under the DFT as well.

\subsection{Computational Details \label{ssec:computational}}

Calculations were performed with a double-$\zeta$ polarized basis set, using norm-conserving
pseudopotentials of the Troullier-Martins type~\cite{pseudos}. Numerical integrals
were carried out on a discretization mesh equivalent to a cutoff of 250 Ry, which
provides total energies for graphene and graphite converged to the few meV range.

\begin{figure}[t!]
  \centering
  \includegraphics[width=1.0\linewidth]{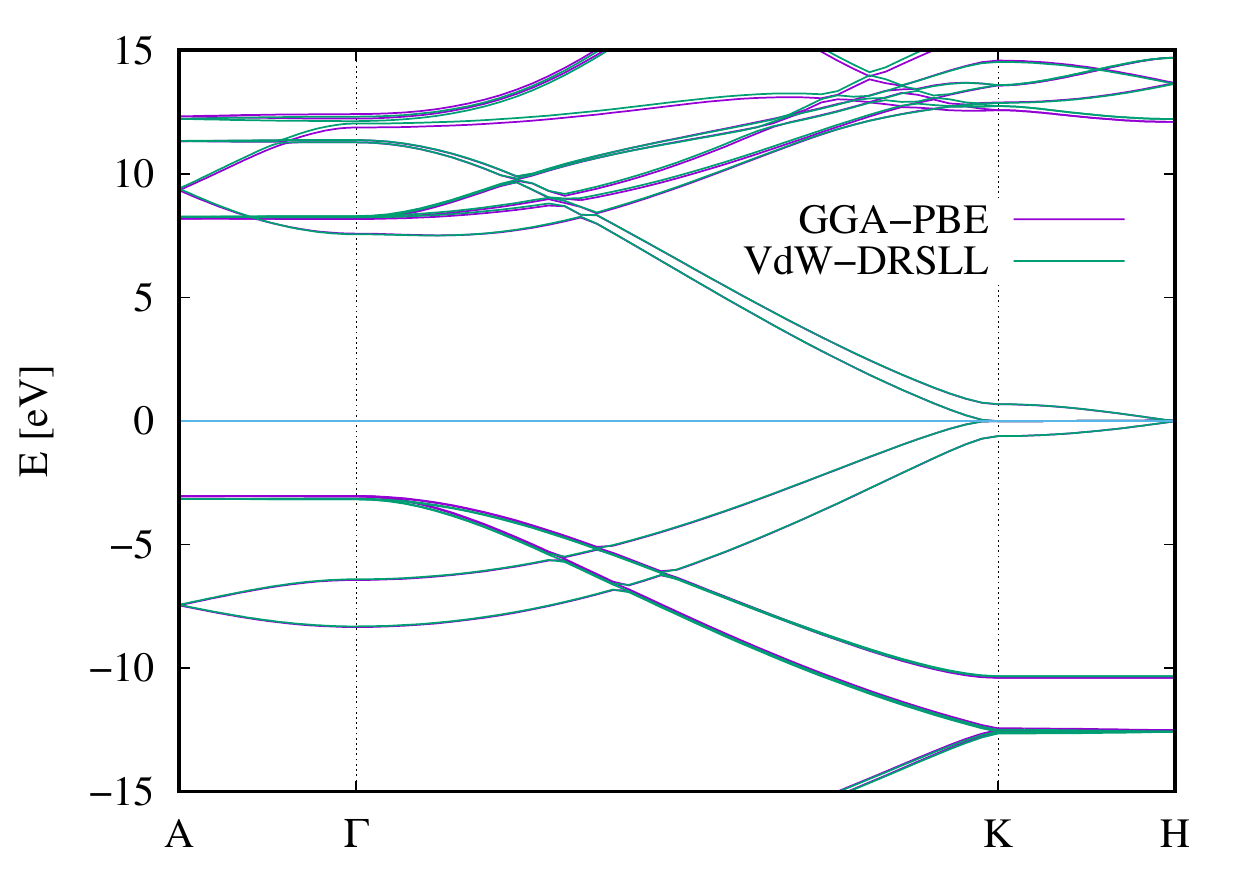}
  \caption{Energy bands for graphite. Purple (green) lines are obtained
  with GGA-PBE (vdW-DRSLL) parametrization, and the structure is taken to be the
  same for both functionals. The clear blue line indicates
  the Fermi Level position.}
  \label{fig:verticalbands}
\end{figure}

The Generalized Gradient Approximation (GGA) in the parametrization
of Perdew-Burke-Ernzerhof (PBE)~\cite{gga-pbe} was used to describe exchange-correlation
effects. GGA accurately describes the lattice parameter
of graphene, but underestimates the interlayer distance \cite{vdw-rdsll}. This, of course,
can be corrected with the use of van der Waals (vdW) type functionals. It has been demonstrated that the
parametrization of Dion-Rydberg-Schr\"oder-Langreth-Lundqvist (DRSLL) of the vdW interaction
provides a good description of the interlayer distance while slightly overestimating
the in-plain lattice constant \cite{vdw-rdsll,vdw-rdsll-2}. Thus, we have carried all
structural relaxations with the vdW-DRSLL functional until residual forces were
less than 0.04~eV/\AA. However, since the GGA is more computationally efficient,
we have used the PBE functional for transport calculations. For a fixed geometry,
the two functionals yield very similar energy dispersions. In \Fref{fig:verticalbands}
the energy bands for bulk graphite are shown, comparing the two functionals.
Around the Fermi level, the energy difference between the two curves is negligible,
and therefore it is concluded that both functionals provide a good description
of the energy of the system. In particular, it must be stressed that the dispersion
along $z$, which is closely related to the interlayer coupling ({\it i.e.}
transport) of the electronic states, is not affected by the passage from the
vdW-DRSLL functional to PBE.

\section{Results \label{sec:results}}

From the {\sc TranSiesta} calculations we obtain the energy-resolved specific
conductance ({\it i.e.} the conductance per unit of tranverse length) for
the different structures.

\begin{figure}[t!]
  \centering
  \includegraphics[width=1.0\linewidth]{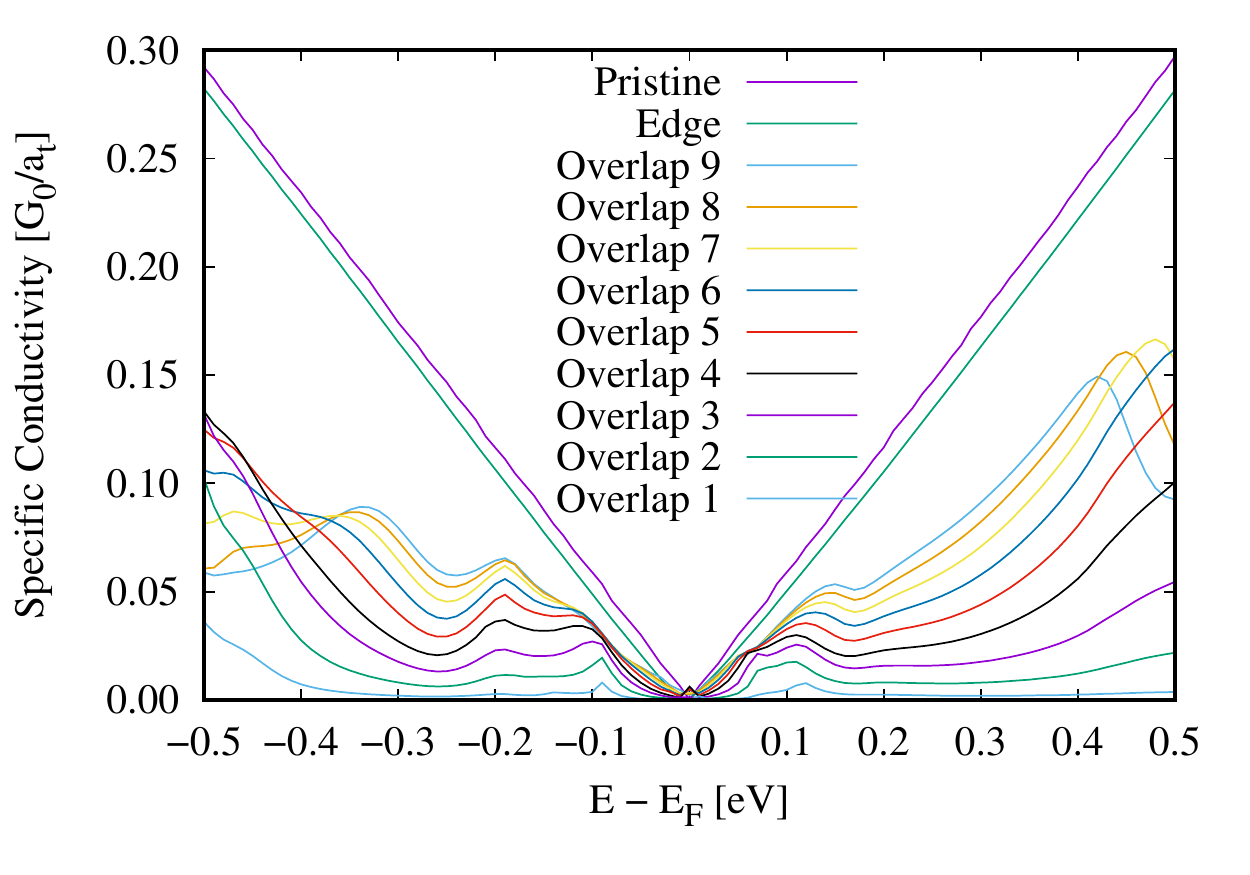}
  \caption{Specific Conductance of the Graphite-Graphene contacts per
  unit of lattice length.}
  \label{fig:conductance}
\end{figure}

In \Fref{fig:conductance} the specific conductance, in units of $G_0= {e^2}/h$ over the
transverse length of the calculation cell ($a_t = 2.484$~\AA), is shown for the
graphite-graphene top contact for different values of the overlap (cf.
\Fref{fig:geometry}) and the edge contact. The pristine graphene case---which
provides the quantum limit for the conductance of the whole structure---
is shown as well for reference. We note that the difference between the edge
contact and the pristine graphene limit is very small, suggesting that an edge,
or large overlap, contact between graphite and graphene would provide a very low
contact resistance.

Regarding the varying amount of overlap, we observe that, contrary to the
metal-graphene case~\cite{xcs_paper}, there is a noticeable monotonic dependence
of the conductance on the overlap width. This is a reflection of the weaker
substrate-graphene $p_z - p_z$ coupling compared to the stronger $d-p_z$ coupling
in metal substrates. Despite the weaker coupling, a relatively narrow overlap of
$\sim 20$~\AA\ suffices to achieve a conductance similar to metallic substrates
(see Ref.~\cite{xcs_paper}).

\begin{figure}[t!]
  \centering
  \includegraphics[width=0.8\linewidth]{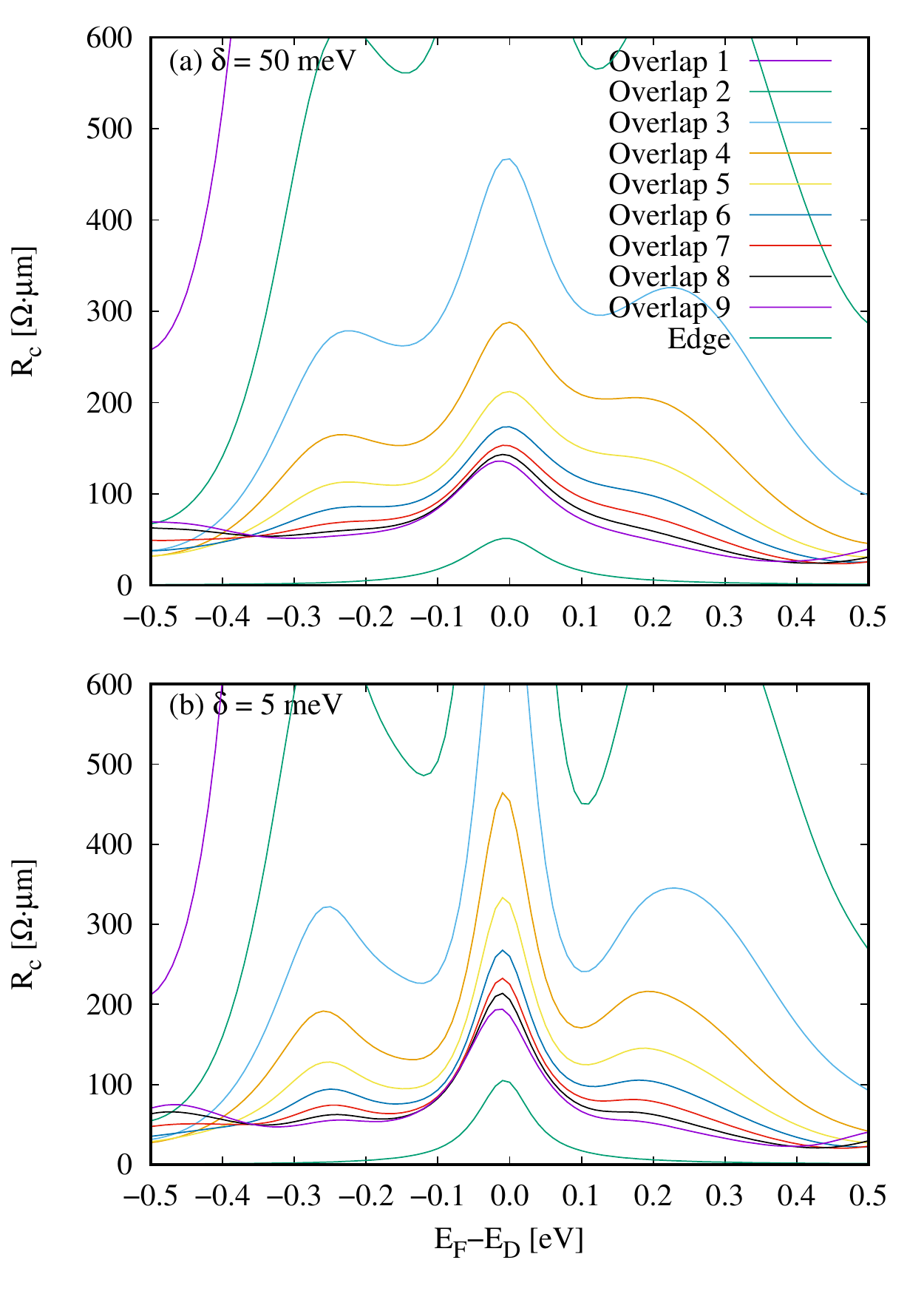}
  \caption{Specific Contact Resistance.
  (a) Thermal broadening at 300K plus electron-hole puddle of 5~meV.
  (b) Thermal broadening at 300K plus electron-hole puddle at 50~meV. }
  \label{fig:contactresistance}
\end{figure}

We now turn our attention to the (specific) contact resistance. It can be
extracted from the calculated conductances of the whole graphite-graphene structure,
$G^{-1}_{gg}(E_{FL}, V)$ and the pristine graphene layer, $G^{-1}_{g}(E_{FL}, V)$:

\begin{eqnarray}
  \fl
  R_c(E_{FL}, V) = G^{-1}_{gg}(E_{FL}, V) - G^{-1}_{g}(E_{FL}, V).
  \label{eq:1}
\end{eqnarray}

Now, in order to calculate the zero bias contact resistance considering a thermal
and gaussian electron-hole (e-h) puddle~\cite{ehpuddles} broadening, we used:

\begin{eqnarray}
  R_c(E_F)= \nonumber \\
  \fl \nonumber
  +~k_B T \left( \int \mkern-12mu \int~\frac{\exp{[(E-E')/k_B T]}}{1+\exp{[(E-E')/k_B T]^2}} \,
  G_{gg}(E)~w(E'-E_F;\eta)~\textrm{d}E~\textrm{d}E' \right)^{-1} \\
  \fl \nonumber
  -~k_B T \left( \int \mkern-12mu \int~\frac{\exp{[(E-E')/k_B T]}}{1+\exp{[(E-E')/k_B T]^2}} \,
  G_{g~}(E)~w(E'-E_F;\eta)~\textrm{d}E~\textrm{d}E' \right)^{-1},
  \label{eq:2}
\end{eqnarray}
where $w(E'-E_F; \eta)$ is the gaussian broadening function and $\eta$ the broadening
paremeter, taken to be $50$~meV for SiO$_2$ substrates~\cite{ehpuddles}.

\begin{figure}[t!]
  \centering
  \includegraphics[width=0.8\linewidth]{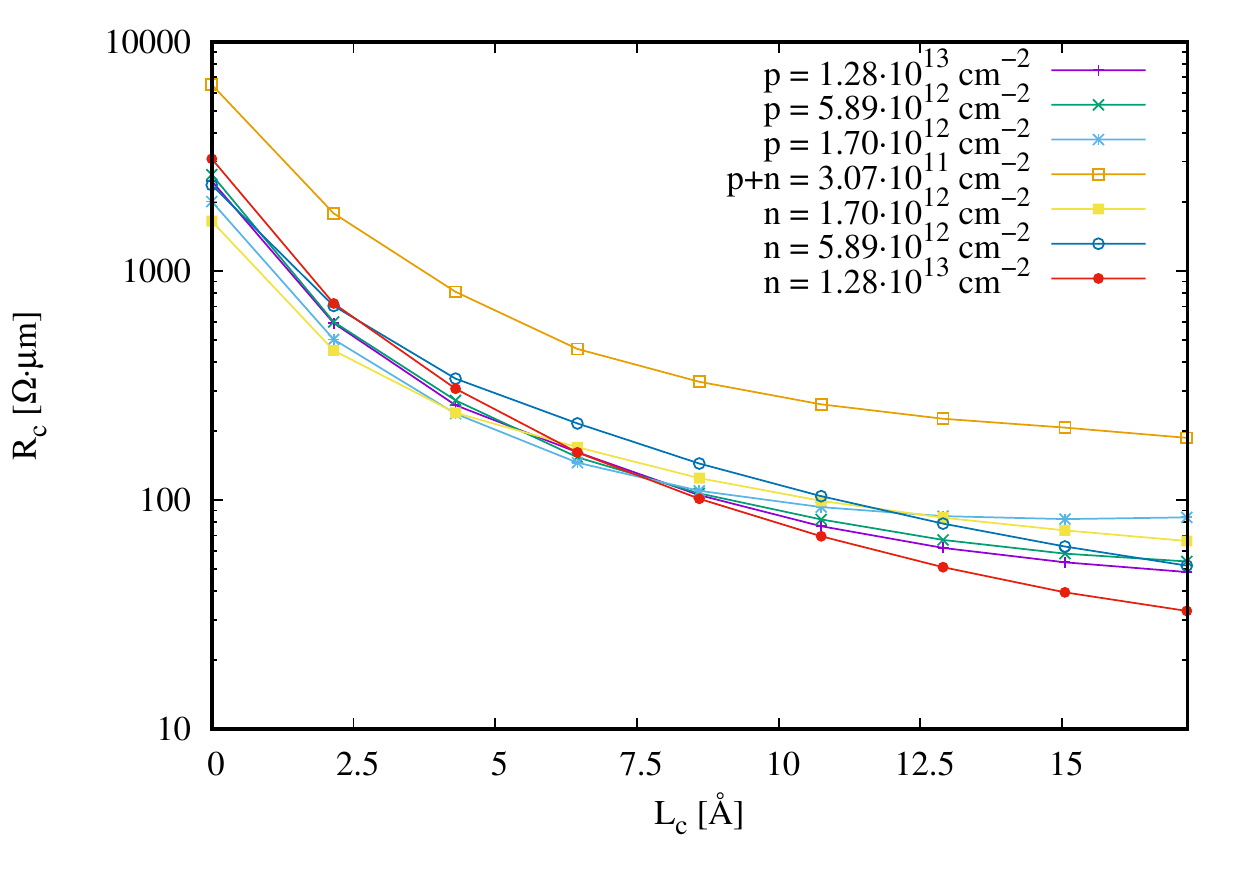}
  \caption{Specific Contact Resistance as a function of contact length for
  different carrier concentrations.}
  \label{fig:Rc_vs_Lc}
\end{figure}

The obtained specific contact resistance results are shown in
\Fref{fig:contactresistance}. The contact resistance at the Dirac point
(undoped graphene) strongly depends on the amount of overlap, with the widest
overlaps getting close to the 100~$\Omega \cdot \mu $m value, especially in the
case of high e-h puddle broadening. The values of the contact
resistance at higher/lower values of the Fermi energy ({\it i.e.} doped samples)
rapidly decrease below the landmark value of 100~$\Omega \cdot \mu $m. These values
are represented, for carrier concentrations, as a function of the
graphene-graphite overlap, $L_c$, in \Fref{fig:Rc_vs_Lc}, where we see that the $R_c$
values effectively saturate for $L_c > \; \sim\!20$~\AA.

\subsection{Current path analysis}

\begin{figure}[t!]
    \centering
    \includegraphics[width=0.7\linewidth]{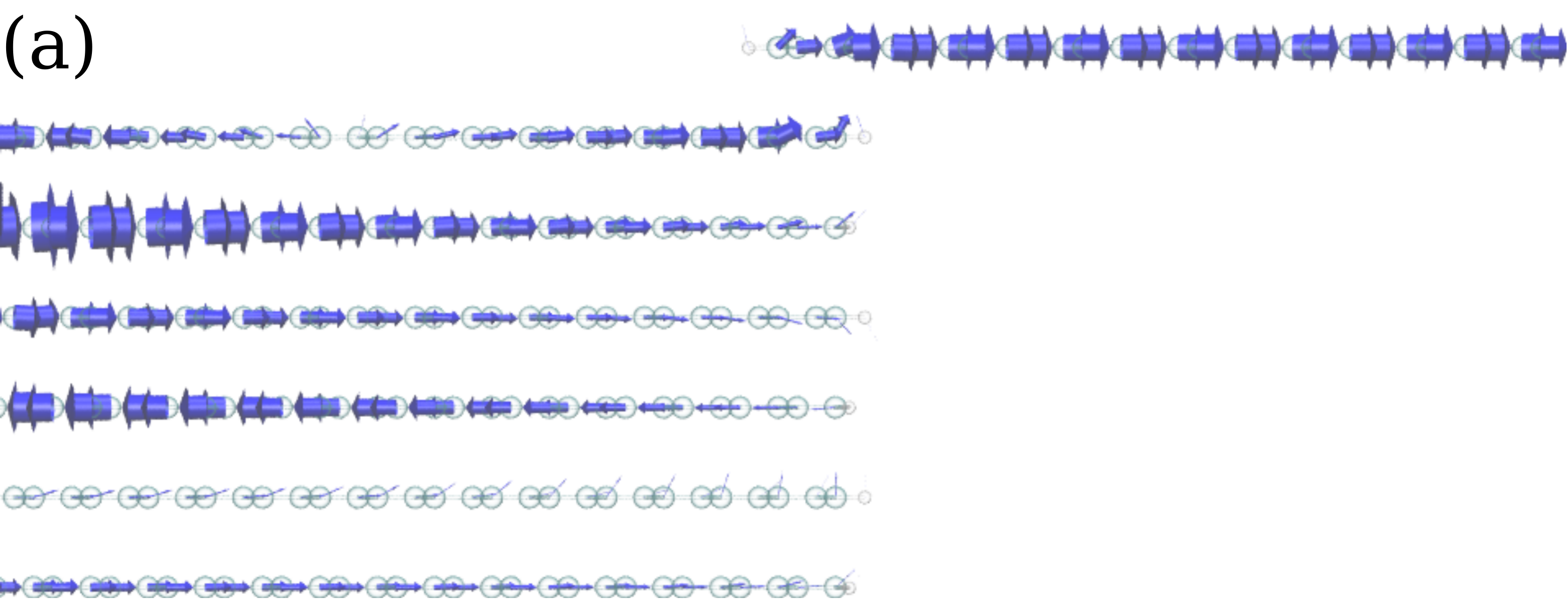} \\
    \vspace{0.5cm}
    \includegraphics[width=0.7\linewidth]{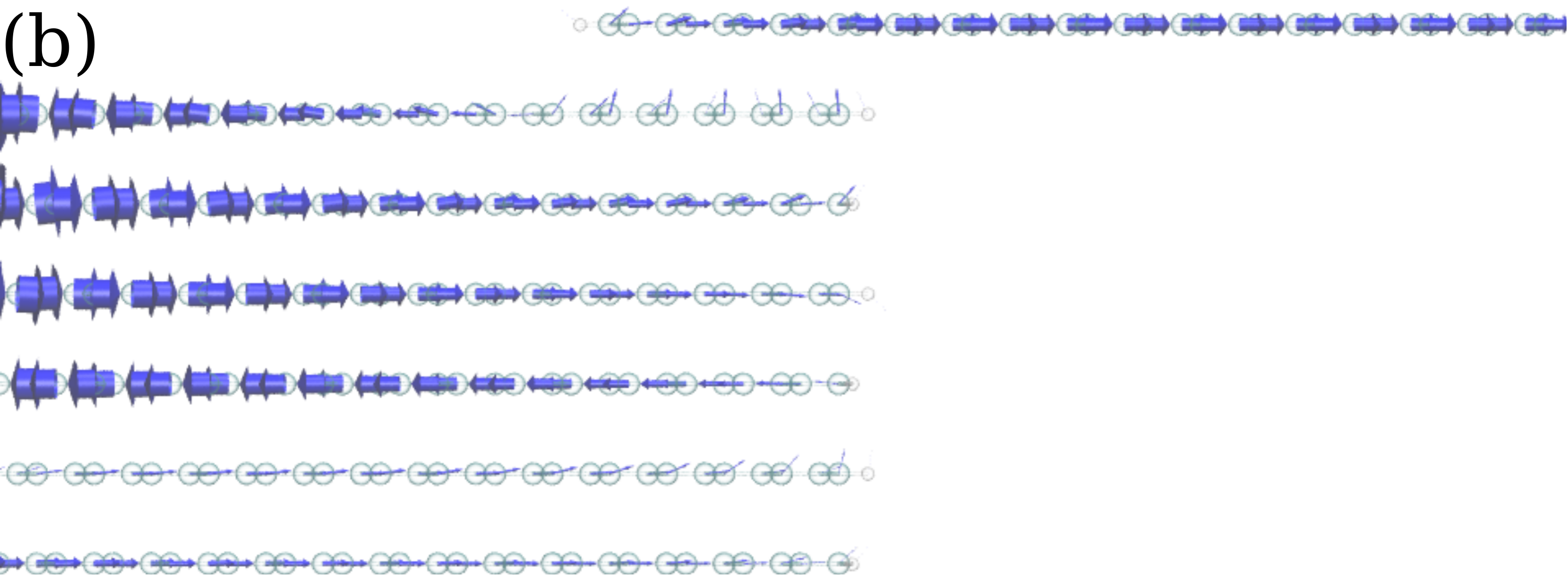} \\
    \vspace{0.5cm}
    \includegraphics[width=0.7\linewidth]{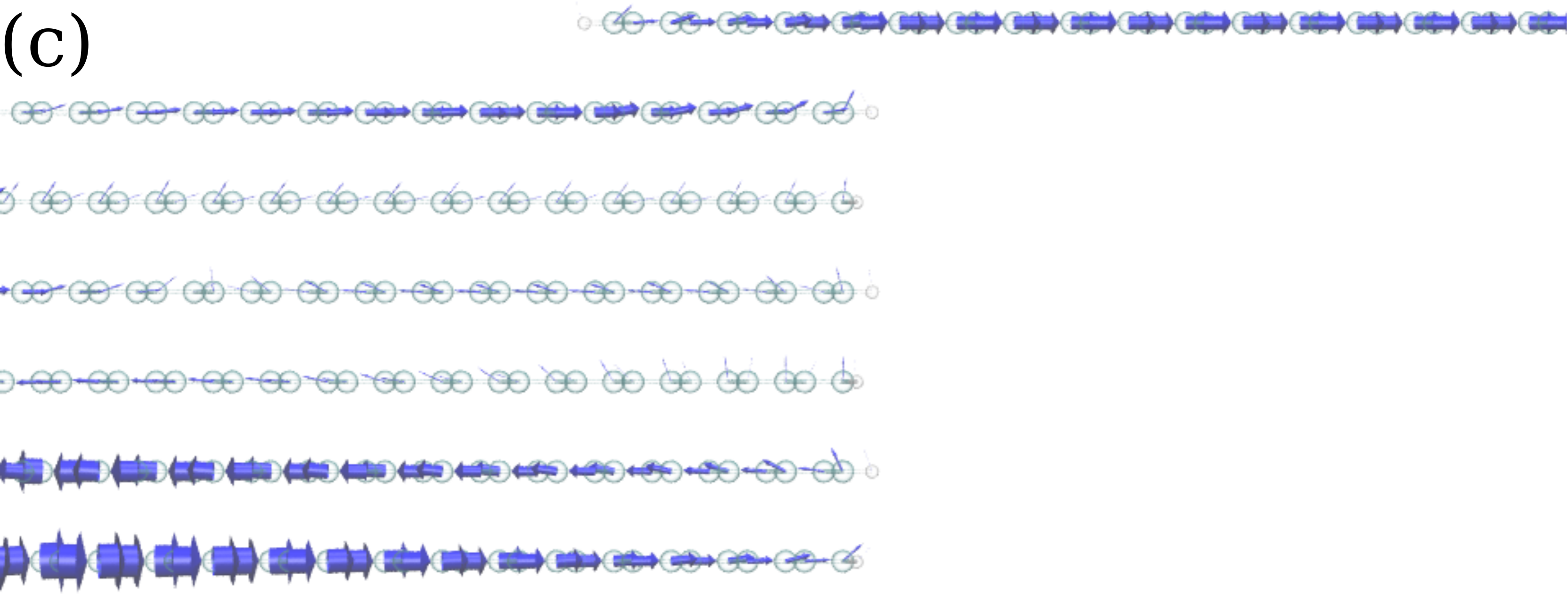} \\
    \vspace{0.5cm}
    \includegraphics[width=0.7\linewidth]{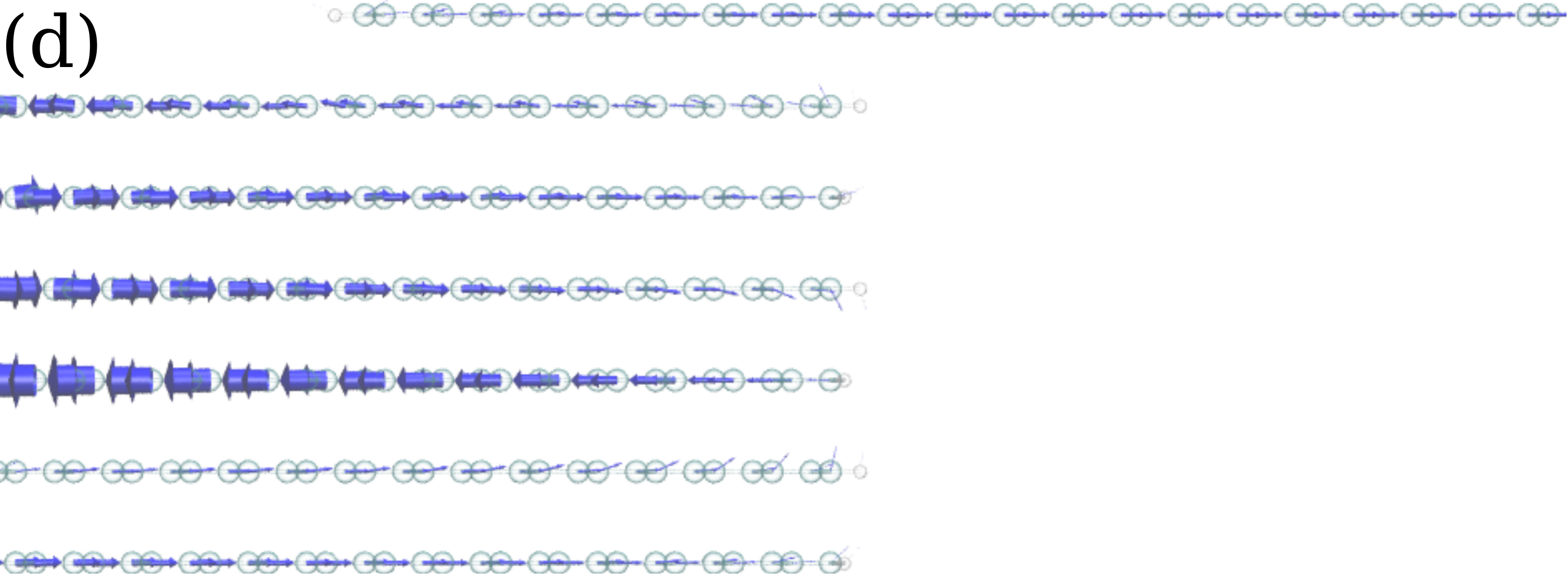} \\
    \vspace{0.1cm}
    \caption{
    Current paths for different contact lenghts for both carriers at fixed $k_{\perp}$ = 0.660~$\pi/a_{\perp}$.\\
    (a) Overlap 2 for e: E =~ 0.105~eV, T = 0.99016 \\
    (b) Overlap 5 for e: E =~ 0.105~eV, T = 0.48904 \\
    (c) Overlap 5 for h: E = -0.105~eV, T = 0.51762 \\
    (d) Overlap 9 for e: E =~ 0.105~eV, T = 0.15161 }
    \label{fig:current}
\end{figure}

In order to asses our conclusions and gain insight into the graphite-graphene coupling, an
eigenchannel analysis has been carried out using the Inelastica package \cite{inelastica}.
In \Fref{fig:current}, current is represented by arrows for each atom in the geometry, represented
by translucid balls, and with the arrow thickness proportional to the magnitude
of the current. Each plot corresponds to a wave vector $k_{\perp}$, perpendicular to the
plane of the representation, different energy of the incoming particle and/or
different overlap, resulting in a transmission coefficient, T.

Figures~\ref{fig:current}.(a)-(d) show the current lines for electrons in the
cases with overlap 2, 5 and 9, with $k_{\perp}$ and $E$ chosen in such a way that
high $T$'s are obtained. We can see that, as the overlap increases, injection
becomes more distributed across the overlapping area, in opposition to the case
of metal-graphene contacts~\cite{xcs_paper}, where only 1-2 metal-carbon bonds
contributed to injection. Notwithstanding that, when only a very small area is
available for injection (e.g. overlap 2), high transmission is still achievable,
with nearly complete injection to graphene taking place through the last two pairs.

\section{Discussion and summary \label{sec:conclusions}}

Of course, any graphite-graphene contact will eventually need to be contacted to
metal leads. One set of measurements of the contact resistance of metal-multilayer
graphene (1,3,4,$\sim$50,$\sim$100 layers) did not find any strong dependence on the
number of layers, which was attributed to only the top layer or two of a graphene
stack playing a role in the contact formation~\cite{metalmultilayer}. It is
expected that a different fabrication procedure promoting the formation of edge
metal-C bonds, such as demonstrated in Ref.~\cite{NiEtchedGContacts}, would
significantly decrease the metal-graphite contact resistance.

In conclusion, we have shown that graphite-graphene contacts provide a promising
route towards the reduction of the contact resistance in graphene FET channels.
Although transfer lengths are significantly higher than in metal-graphene
contacts, their magnitudes are still quite small, at $\sim$20~\AA.
In addition, edge graphite-graphene contacts are expected to have exceptionally
low contact resistance.

\ack
We acknowledge financial support by the Spanish MINECO
under Project No. TEC2015-67462-C2-1-R (MINECO/
FEDER). Also, this project has received funding from the
EU Horizon2020 research and innovation program under
grant No. 696656, Graphene Flagship and under grant No.
GrapheneCore2 785219.
\section*{References}
\bibliographystyle{iopart-num}
\bibliography{paper}

\end{document}